\documentclass[aps,showpacs,floats]{revtex4}

\usepackage{epsfig}

\newlength{\mylen}

\begin{document}

\title{Effective boost and ``point-form'' approach}

\author{B. Desplanques}
\email{desplanq@isn.in2p3.fr}
\author{L. Theu{\ss}l}
\email{lukas@isn.in2p3.fr}
\affiliation{Institut des Sciences Nucl\'eaires,  
 F-38026 Grenoble Cedex, France }

\author{S. Noguera}
\email{Santiago.Noguera@uv.es}
\affiliation{Departamento de Fisica Teorica, 
Universidad de Valencia, E-46100 Burjassot, Spain}

\begin{abstract}
Triangle Feynman diagrams can be considered as describing form 
factors of  states bound by a zero-range interaction. These form 
factors are calculated for scalar particles  and compared to  
point-form and non-relativistic results. 
By examining the expressions of the complete calculation in different frames, 
we obtain an effective boost transformation which can be compared to the 
relativistic kinematical one underlying the present point-form calculations, 
as well as to the Galilean boost. 
The analytic expressions obtained in this simple model allow a
qualitative check of certain results obtained in similar studies. In
particular, a mismatch is pointed out between recent practical applications of 
the point-form approach and the one originally proposed by Dirac.
\end{abstract}

\maketitle

\noindent
The point-form calculation of form factors in impulse approximation has 
recently been performed for various systems: the deuteron~\cite{KLIN0}, the 
nucleon~\cite{WAGE} and a bound state of two scalar particles interacting via 
the exchange of a massless scalar boson~\cite{DESP0}. With respect to a 
non-relativistic calculation, discrepancies with experimental data are slightly
increased
in the first case and essentially removed in the second one. For the 
last case, there is no experiment but a more complete calculation in ladder 
approximation can be performed. Indeed, the Bethe-Salpeter 
equation~\cite{BETH} can then be solved easily, due to a hidden symmetry 
in the Wick-Cutkosky model~\cite{WICK,CUTK}. Moreover, this model accounts for 
minimal features of realistic interaction models, such as 
the field-theory character or explicit relativistic covariance.
For this system, the exact and point-form calculations of the 
form factors strongly differ~\cite{DESP0}, especially when the 
zero-mass limit is approached or when large $Q^2$ are considered, two limits 
that a relativistic approach should be able to deal with. This test case points 
to an important contribution from two-body currents. Amazingly, a 
calculation using the non-relativistic expressions for the form factors with the 
point-form wave functions does rather well. The difference resides mainly in 
the expression of the transformed momenta under a boost. Unfortunately, 
the complexity of the calculation in the Wick-Cutkosky model prevents one 
from checking this point directly on the expression of the form factors.

However, there exists a simpler model which allows for a completely analytic 
solution and
which therefore is particularly useful to gain insight into the
origin of the above mentioned properties. This model describes a bound state of
two scalar particles together with a zero-range interaction, providing another 
case where the Bethe-Salpeter equation may be solved in closed form. 
Contrary to the Wick-Cutkosky model however, the interaction now is of short 
range, which allows one to check qualitative features obtained in 
ref.~\cite{DESP0} with a different interaction model.
The fully covariant amplitude for the bound state of total mass $M^2=P^2$ in this 
case is given by:
\begin{equation}
\label{bsam}
\Phi (P,p) = \frac{C}{\left(m^2-p^2\right)\left(m^2-\big(P-p\big)^2\right)},
\end{equation}
where $p$ is the four-momentum of one of the two
constituents with equal mass $m$, and $C$ is a normalization constant.

\begin{figure}[htb]
\begin{center}
\mbox{\psfig{file=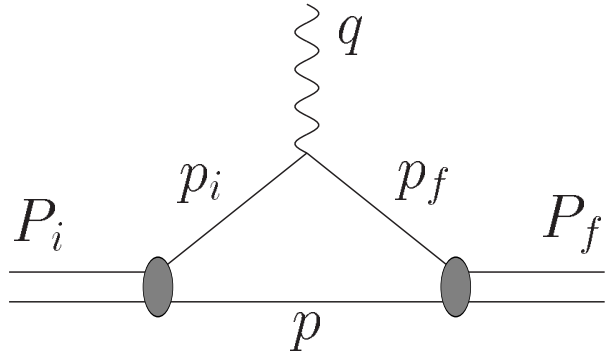,width=8cm}}
\end{center}
\caption{Representation of a virtual photon absorption on a two-body system 
with the kinematical definitions.}\label{fig1}
\end{figure}

This solution may be used to calculate form factors of the bound state in closed
form. These form factors are represented by the triangle Feynman diagram shown 
in Fig.~\ref{fig1}. Of particular interest in the present context is the 
expression obtained for the electromagnetic form factor 
after integrating over the zero component of the spectator particle four
momentum. In the Breit frame it reads:
\begin{equation}
\label{breit}
F_1(Q^2)= \frac{C^2}{2} \int \frac{d\,\vec{p}}{(2\,\pi)^3} \;
\frac{(e_i+e_f+2\,e_p)}{e_p\,(e_i+e_f)\,(E^2-(e_i+e_p)^2)\,(E^2-(e_f+e_p)^2)},
\end{equation} 
with $E=\sqrt{M^2+(\vec{Q}/2)^2}$, $e_i =\sqrt{m^2+(\vec{p}+\vec{Q}/2)^2}$,
$e_f=\sqrt{m^2+(\vec{p}-\vec{Q}/2)^2}$ and $e_p=\sqrt{m^2+\vec{p}^{\;2}}$. 

\noindent
In this form it does not contain an obvious 
track of a bound state but this feature becomes apparent when looking at the 
expression at $q^{\mu}=0$: 
\begin{equation}
F_1(0) = \frac{C^2}{2} \frac{1}{M}\, \int \frac{d \vec{k}}{(2\,\pi)^3} \left[
\;\left(\frac{1}{2\,e_k\,(M-2\,e_k)}\right)^2 
-\left(\frac{1}{2\,e_k\,(M+2\,e_k)}\right)^2 \right].
\label{bo4}
\end{equation}
The dominant first term of this expression
is precisely what we get for the norm of a system described by the wave 
function $\phi(\vec{k})=c/[2\,e_k\,(M-2\,e_k)]$, which itself corresponds 
to a zero-range interaction in a semi-relativistic potential model. 
This wave function has the structure expected for a solution of the mass 
operator, $M=2\,e_k+V$ (in the c.m.). Therefore, we have another model where 
the point-form approach in impulse approximation can be checked reliably 
since the calculation of the Feynman diagram of Fig.~\ref{fig1} is 
straightforward.

Results for the elastic charge form factor in the point-form 
approach can be obtained from expressions given in ref.~\cite{DESP0}: 
\begin{equation}
\label{ptf}
F_1(Q^2)= \frac{1+v^2}{\sqrt{1-v^2}} \int \frac{d \vec{p}}{(2\pi)^3} 
\;\phi_f(\vec{p}_{tf}) \;\phi_i(\vec{p}_{ti}),\label{b10}
\end{equation}
where the velocity $\vec{v}$ is related to the momentum transfer by the 
relation $v^2=Q^2/(Q^2+4M^2)$ for the elastic case, and
the Lorentz-transformed momenta are defined as:
$(p^x,p^y,p^z)_{tf}= (p^x,p^y, \frac{p^z-v \, e_p}{\sqrt{1-v^2}})$ and
$(p^x,p^y,p^z)_{ti}= (p^x,p^y, \frac{p^z+v \, e_p}{\sqrt{1-v^2}})$. 
A similar relation holds for the scalar form factor~\cite{DESP2}.

Results of a non-relativistic calculation, 
\begin{equation}
\label{nr}
F_1(Q^2)=  \int \frac{d \vec{p}}{(2\pi)^3} \;\;
\phi_f\Big(\vec{p}-\frac{1}{4}\vec{q}\,\Big)
\;\; \phi_i\Big(\vec{p}+\frac{1}{4}\vec{q}\,\Big),
\end{equation}
which implies the use of the same wave functions $\phi(\vec{p}\,)$ as previously 
quoted together with a Galilean boost, will also be considered.
This is mainly done to emphasize the role of the different boost expressions, as
discussed below, the relevance of a non-relativistic calculation being a priori 
questionable for the high momentum transfers (up to $Q^2= 100\,m^2$) that we
will consider.

\begin{table}[htb]
\setlength\arraycolsep{10pt}
\caption{Elastic scalar- and vector-  form factors, $F_0(q^2)$ and  $F_1(q^2)$.
Results are given successively for the Feynman diagram
($F.D.$) (identical to a Bethe-Salpeter calculation with a zero-range
interaction), the point-form approach ($P.F.$) and the non-relativistic case
($N.R.$), and for two total masses.}\label{t10}
\begin{center}
\begin{ruledtabular}
\begin{tabular}{lccccc}
$Q^2/m^2$	   & 0.01  &  0.1    &  1.0    &  10.0    & 100.0   \\  \hline
$M=1.6\,m$ \rule{0pt}{3ex} & &       &         &	  &	    \\
$F_0\;\;\;\;F.D.$  & 1.325 & 1.309   & 1.176   & 0.659    & 0.187   \\ [0.ex]
$F_1\;\;\;\;F.D.$  & 0.999 & 0.989   & 0.908   & 0.566    & 0.191   \\ [0.ex]
$F_0\;\;\;\;P.F.$  & 0.511 & 0.499   & 0.393   & 0.095    & 0.33-02 \\ [0.ex]
$F_1\;\;\;\;P.F.$  & 0.998 & 0.989   & 0.904   & 0.485    & 0.82-01 \\ [0.ex]
$F_{0,1}\;\;N.R.$  & 0.998 & 0.991   & 0.923   & 0.595    & 0.210   \\ [1.ex]
$M=0.1\,m$	   &	   &	     &         &	  &	    \\
$F_0\;\;\;\;F.D.$  & 1.498 & 1.487   & 1.389   &  0.920   & 0.320   \\ [0.ex]
$F_1\;\;\;\;F.D.$  & 0.999 & 0.995   & 0.954   &  0.723   & 0.315   \\ [0.ex]
$F_0\;\;\;\;P.F.$  & 0.015 & 0.27-02 & 0.87-04 &  0.15-05 & 0.20-07 \\ [0.ex]
$F_1\;\;\;\;P.F.$  & 0.956 & 0.482   & 0.74-01 &  0.77-02 & 0.78-03 \\ [0.ex]
$F_{0,1}\;\;N.R.$  & 0.999 & 0.998   & 0.979   &  0.840   & 0.465   \\ [1.ex]
\end{tabular}
\end{ruledtabular}
\end{center}
\end{table}

A sample of results is given in Table \ref{t10} for both a scalar and a vector 
(charge) form factor, and for two masses, $M=1.6\,m$ and an extreme 
relativistic case, $M=0.1\,m$.
One notices that the point-form results fail to
reproduce the exact ones, especially in the limit $M \rightarrow 0$ or for the
scalar probe. They qualitatively confirm results obtained in ref.~\cite{DESP0}, 
despite a significant difference in the interaction models:
long-range in ref.~\cite{DESP0}, short range in the present work. 
The fact that the non-relativistic
calculation does well in all cases strongly suggests that the effective
boost  is closer to the non-relativistic than to the kinematical relativistic
one. Some of the details may be discussed, but the size of the discrepancy is so
large, especially when the total mass M goes to zero or $Q^2$ goes to $\infty$,
that there is no doubt on the diagnostic. We remark that there also exist
discrepancies for $F_0$ at low $Q^2$. They may be accounted for by two-body 
currents that have a close relationship to Z-type diagrams in an instant-form
approach. They are not considered here since they are
common to both the point-form and the non-relativistic calculations.

We shall now use the different expressions obtained above to directly 
compare the boost transformations applied in different approaches. This will
provide further insight into the features evidenced by the results of
Table~\ref{t10} and those presented in refs.~\cite{KLIN0,WAGE,DESP0}. 
The transformed momenta,
when a system of total mass $M$ is boosted to acquire a momentum  
$\vec{P}$ (equal to $\vec{q}/2$ in the Breit frame employed in 
Eqs.~(\ref{ptf},\ref{nr})), in the Galilean-covariant and the point-form 
approach, are respectively given by:
{
\settowidth{\mylen}{$\displaystyle\vec{p}=\displaystyle\vec{k}+
\displaystyle\frac{\vec{k}  \cdot  \vec{P} \;\, 
\vec{P}}{\vec{P}^2}\left(\sqrt{1+\vec{P}^2/M^2} \;\; - 1\right) +
\displaystyle\frac{\vec{P}}{2}\; \displaystyle\frac{2\sqrt{m^2+\vec{k}^2}}{M},$} 
\begin{eqnarray}
\label{bo1}
&\makebox[\mylen][t]{$\displaystyle\vec{p}= 
\displaystyle\vec{k}+\displaystyle\frac{\vec{P}}{2},$}
\qquad&({\rm non-relativistic})  \\
\label{bo2}
&\displaystyle\vec{p}=\displaystyle\vec{k}+
\displaystyle\frac{\vec{k}  \cdot  \vec{P} \;\, 
\vec{P}}{\vec{P}^2}\left(\sqrt{1+\vec{P}^2/M^2} \;\; - 1\right) +
\displaystyle\frac{\vec{P}}{2}\; \displaystyle\frac{2\sqrt{m^2+\vec{k}^2}}{M},
\qquad&({\rm point-form})
\end{eqnarray}}

\noindent
where  $\vec{p}$ and $\vec{k}$ are the momenta of the spectator
particle in the moving and in the c.m.  
frames. A large part of the difference between non-relativistic and point-form
results is due to the presence of the factor 
$2\sqrt{m^2+\vec{k}^2}/M$ multiplying $\vec{P}/2$ in Eq.~(\ref{bo2}).  It
tends to make the momentum transfer actually larger than what it should be, 
hence larger charge radii and faster fall-off of form factors as it is
observed from results presented in refs.~\cite{WAGE,DESP0}. The fact that the 
non-relativistic calculation does relatively well suggests that this factor is 
effectively absent in the exact calculation of form factors. 

A qualitative confirmation is obtained by the examination of Eq.~(\ref{breit}), 
which indicates that  the 
combination of the vectors $\vec{p}$ and $\vec{Q}/2$, that appears in many 
places, is the same as given by a Galilean boost, Eq.~(\ref{bo1}). This feature 
has a general character and it
explains why a non-relativistic calculation of form factors is not so 
bad, contrary to what might be expected a priori. Similar results hold for the 
form factor relative to a scalar probe.

We first analyze the expression of the exact form factors, Eq.~(\ref{breit}), 
at $q^{\mu}=0$. In this case, the calculation 
can be performed by integrating over the time component of the spectator 
particle, for the system at rest or with arbitrary momentum $\vec{P}$. As the 
result is a covariant one, it should be the same in any frame, thus providing a 
relation between the momenta of the spectator particle in the c.m. and in the 
moving system. The result is given by:
\begin{equation}
\vec{p}= \vec{k} +\frac{\vec{k}  \cdot \hspace{-1mm} \vec{P} \;\, 
\vec{P}}{\vec{P}^2}\Big( \sqrt{ 1+\vec{P}^2/(4\,e_k^2) } \;\; - 1 \Big) 
+\frac{\vec{P}}{2}. 
\label{bo5}
\end{equation}
This relation, which has been mentioned in a slightly different 
context in ref.~\cite{HAMM}, corresponds to an instant-form approach. 
Even though it does not exhibit an explicit dependence on the interaction, as
it is the case for the transformation of Eq.~(\ref{bo2}), one has to remember
that the relevant boost parameter is the velocity 
$\vec{v}/\sqrt{1-v^2}=\vec{P}/M$. Expressing
Eqs.~(\ref{bo2},\ref{bo5}) in terms of this quantity shows the interaction and
kinematics dependence of the boost in instant form and in point form, 
respectively, as it should be.

While Eq.~(\ref{bo5}) provides the desired result (absence of the factor $2\,e_k/M$ in 
front of $\vec{P}$), it is too early to conclude at this point. Indeed, other 
expressions of  Eq.~(\ref{breit}) can be obtained by performing 
the calculation of the Feynman triangle diagram, Fig.~\ref{fig1}, in different 
ways. Instead of integrating over the time component of the four-momentum $p$, 
one can integrate over any linear combination of the four components, 
$\lambda \cdot p$, where $\lambda^2=1$. This 
would correspond to a description on a hyperplane 
$\lambda \cdot x=ct$. As the form factor is invariant under a  Lorentz 
transformation and since
the above change in the integration variable amounts to such a 
transformation with velocity $\vec{v}=\vec{\lambda}/\lambda_0$, a 
generalized relation immediately follows from Eq.~(\ref{bo5}):
\begin{equation}
\vec{p}=\vec{k}+ 
\vec{\lambda}\,\left(\frac{\vec{k}\cdot\vec{\lambda}}{\lambda_0+1} 
+\frac{\vec{k}\cdot \vec{P'}}{2\,e_k }\right)
+\frac{\vec{k}\cdot \vec{P'} \; \Big(\vec{P'}+\vec{\lambda}\;
\vec{\lambda}\cdot \vec{P'}/(\lambda_0+1)\Big)}
{4\,e_k^2\;\Big(1+\sqrt{1+\vec{P'}^2/(4\,e_k^2)}\;\Big) }
+\frac{\vec{P}}{2}\; 
+\frac{\vec{\lambda}}{2} \; \left( \sqrt{4\,e_k^2+\vec{P'}^2}- 
\sqrt{M^2+\vec{P'}^2} \right), 
\label{bo6}
\end{equation}
where $\vec{P'}$ is related to $\vec{P}$ by the Lorentz transformation:
\begin{equation}
\label{new}
\vec{P'} =\vec{P}+ \vec{\lambda}\,
\frac{\vec{\lambda} \cdot \vec{P}}{\lambda_0+1}-\vec{\lambda}\, E_P,
\end{equation}
with $E_P=\sqrt{M^2+\vec{P}^2}$.
For our purpose, only the two  terms in Eq.~(\ref{bo6}) that do not vanish in 
the limit $\vec{k} \rightarrow 0$ are relevant. 
As it can be observed, the bracket in the very last term is proportional to
$(2\,e_k-M)$. This factor corresponds to an effective interaction term and 
is characteristic of the contribution that appears here or there, 
depending on the relativistic approach.
Thus, despite apparent differences, none of the expressions given by
Eqs.~(\ref{bo6},\ref{new}) is favoured in calculating form factors, 
provided evidently that two-body currents 
originating from Z-type diagrams are included. In particular, it can be shown 
that Eq.~(\ref{bo6}) allows one to recover the relation used in point-form 
calculations, Eq.~(\ref{bo2}), by choosing $\lambda_0=E_P/M, 
\;\vec{\lambda}=\vec{P}/M$ (implying $\vec{P'}=0$).

The above results can be extended to the case $q^{\mu} \neq 0$. While doing so, 
one has to take care that the initial and final states are described on the 
same
surface, $\lambda \cdot x = ct$. Examining the recent point-form calculations, 
it is obvious that this condition is not fulfilled. The initial and final 
states are obtained from the state at rest, which is described on a 
surface $\lambda \cdot x = ct$, with $\lambda=(1,0,0,0)$. Applying 
a kinematical boost amounts to using a relation such as Eq.~(\ref{bo6}), 
with the appropriate $\lambda$. Thus, the states required for an estimate of 
form factors, in the Breit frame for instance, correspond to a description on 
two different surfaces,  $\lambda \cdot 
x = ct$ with $(\lambda_0, \; \vec{\lambda})=(1, \; -\vec{v})/\sqrt{1-v^2}$ for 
the initial state and $(\lambda_0,\; \vec{\lambda})=(1, \; 
\vec{v})/\sqrt{1-v^2}$ for the final one. This is not the standard way of
performing calculations that, generally, refer to one and the same definition 
of the surface on which quantization is performed, independently of the 
formalism used to implement relativity. 

Having emphasized a possible mismatch  in the description of initial and final 
states in the point-form calculations of form factors made up to
now~\cite{KLIN0,WAGE}, we go 
back to the various expressions of the transformation of the momentum under a 
boost, Eqs.~(\ref{bo1} -- \ref{bo6}). From the above 
discussion, the relevant boost effect is determined by the 
relative boost of the initial and final states, both being described on the 
same surface, like the one given in Eq.~(\ref{bo5}). Taking into account that 
$\vec{P}$ is directly associated to the momentum transfer, one can anticipate 
that the effective enhancement of the momentum transfer due to the factor 
$2\,e_k/M$ in Eq.~(\ref{bo2}), which is responsible for the largest effect 
obtained in the point-form calculations of form factors, is essentially absent, 
as it can be checked on the full expression of the form factor of the system 
considered here. Unfortunately, an examination of Eq.~(\ref{breit}) does not
provide an effective boost transformation as it was the case before, however, it
is clear that there do not arise any additional factors $2e_k/M$, suggesting 
that this effective boost is still closer to the non-relativistic than to the
kinematical relativistic one.

In this paper, we considered another example where a comparison of a point-form 
calculation of form factors with an exact one is possible. It confirms earlier 
conclusions while providing analytical checks~\cite{DESP0}. It is shown 
that the applications of the point-form approach made until 
now~\cite{KLIN0,WAGE} imply the 
description of the initial and final states on two different surfaces. This 
prevents one from considering time-ordered contributions 
(the invariant times for the two states are not defined in the same way). 
Interestingly, the effective boost that allows one to construct the final state 
from the initial one is closer to the non-relativistic than to the kinematical 
relativistic one, explaining the relative success of a non-relativistic 
description of form factors in a domain where it is not expected. Not 
surprisingly, the difference involves terms proportional to $2\,e_k-M$, which 
can be turned into an interaction term. The effective boost transformation 
accounts for a contribution originating from the free-particle energy, included 
in the present point-form calculation, but also one from the interaction 
term, which has to be accounted for separately in the point-form approach. 
When the system 
under consideration is given a velocity $\vec{v}$ (momentum $\vec{P}$), the two 
contributions are roughly given by $\simeq\vec{v}\,(2\,e_k)/\sqrt{1-v^2}$ 
(which represents a  contribution to the momentum enhanced by a factor
$2\,e_k/M$) and $\simeq\vec{v}\,(M-2\,e_k)/\sqrt{1-v^2}$ 
(which cancels this enhancement). We recall that this last term, whose 
corresponding contribution is obviously absent in the present point-form 
calculations of form factors, can be cast into 
an interaction term by using the expression of the mass operator, $2e_k+V=M$.

The failure of the currently applied point-form approach can be remedied by 
brute force, by adding two-body currents that are constructed such that they 
reproduce what is 
expected from Feynman diagrams once the one-body current is accounted for. 
One can hope to fulfill in this way current 
conservation and reproduce the Born amplitude, two constraints that are relevant 
at small and large momentum transfers, respectively. Some work along these 
lines is in progress~\cite{DESP2}.

It is noticed that the present point-form calculations rely in practice 
on the use of wave functions obtained from quantization on  a hyperplane 
$\lambda \cdot x = ct$. Originally, in Dirac's work~\cite{DIRA}, this 
operation was supposed to be performed on a hyperbolo\"{\i}d, $x \cdot x = ct$. 
One can imagine that an approach fully consistent with this feature could 
remove some of the encountered problems. Despite a few 
attempts~\cite{FUBI,GROM}, not much has been done along these lines however.

An intermediate approach, suggested by the above analysis,  is to change the 
quantization surface for the initial or for the final state, 
or both, so that they coincide after a boost from the state at rest.
This involves the dynamics, with the result that the original 
simplicity of the approach is lost. 
This procedure could be realized in an instant-form 
calculation, but the answer is not unique. However, starting from the 
Breit-frame results, covariant ones are obtained by the particular choice, 
$\lambda^{\mu}=(P_i+P_f)^{\mu}/\sqrt{(P_i+P_f)^2}$, the only one consistent 
with the symmetry between initial and final states. With this definition, the 
quantization surface, $\lambda \cdot x= ct$, is unchanged under a Lorentz
transformation  while 
it is modified in a standard instant-form calculation where $\lambda=(1,0,0,0)$. 

This invariance is not equivalent to the one resulting from the consideration of
a hyperbolo\"{\i}d surface. Nevertheless, by proceeding this way one recovers 
some of the 
properties expected in a full point-form calculation. From what we checked, the 
major problems evidenced by the present point-form calculations are removed. 
It has to be shown however, whether a fully consistent approach can be 
constructed along these lines.

\bigskip
{\bf Acknowledgements}
This work has been supported in part by the EC-IHP Network ESOP, contract 
HPRN-CT-2000-00130 and the DGESIC (Spain), under contract PB97-1401-C02-01.


\begin{thebibliography}{99} 
\bibitem{KLIN0} T.W. Allen, W.H. Klink and W.N. Polyzou, Phys. Rev. C 63, 034002 
(2001).
\bibitem{WAGE} R. Wagenbrunn et al.,  Phys. Lett. B 511, 33 (2001) .
\bibitem{DESP0} B. Desplanques and L. Theu{\ss}l, nucl-th/0102060.
\bibitem{BETH} E.E. Salpeter and H.A. Bethe, Phys. Rev.  84, 1232 (1951).
\bibitem{WICK} G.C. Wick, Phys. Rev.  96, 1124 (1954).
\bibitem{CUTK} R.E. Cutkosky, Phys. Rev.  96, 1135 (1954).
\bibitem{DESP2} B. Desplanques and L. Theu{\ss}l, in preparation.
\bibitem{HAMM} B. Hamme and W. Gl\"ockle, Few-Body Systems 13, 1 (1992).
\bibitem{DIRA} P.A.M. Dirac, Rev. Mod. Phys. 21, 392 (1949). 
\bibitem{FUBI} S. Fubini, A.J. Hanson and R. Jackiw, Phys. Rev. 7, 1732 (1973).
\bibitem{GROM} D. Gromes, H. J. Rothe and B. Stech, Nucl. Phys. B75, 313 (1974).
\end{thebibliography}
\end{document}